\documentclass[twocolumn]{aastex63}

\graphicspath{{./}{figures/}}

\shorttitle{Thermal and gravitational energy densities of the LSS}
\shortauthors{Chiang et al.}

\begin{document}
\title{The thermal and gravitational energy densities in the large-scale structure of the Universe}

\author[0000-0001-6320-261X]{Yi-Kuan Chiang}
\affiliation{Center for Cosmology and AstroParticle Physics (CCAPP), The Ohio State University, Columbus, OH 43210, USA}

\author[0000-0001-5133-3655]{Ryu Makiya}
\affiliation{Kavli Institute for the Physics and Mathematics of the Universe (Kavli IPMU, WPI), University of Tokyo, Chiba 277-8582, Japan}

\author[0000-0002-0136-2404]{Eiichiro Komatsu}
\affiliation{Max-Planck-Institut f\"{u}r Astrophysik, Karl-Schwarzschild Str. 1, 85741 Garching, Germany}
\affiliation{Kavli Institute for the Physics and Mathematics of the Universe (Kavli IPMU, WPI), University of Tokyo, Chiba 277-8582, Japan}

\author[0000-0003-3164-6974]{Brice M\'enard}
\affiliation{Department of Physics \& Astronomy, Johns Hopkins University, 3400 N. Charles Street, Baltimore, MD 21218, USA}
\affiliation{Kavli Institute for the Physics and Mathematics of the Universe (Kavli IPMU, WPI), University of Tokyo, Chiba 277-8582, Japan}

\begin{abstract}
As cosmic structures form, matter density fluctuations collapse gravitationally and baryonic matter is shock-heated and thermalized. We therefore expect a connection between the mean gravitational potential energy density of collapsed halos, $\Omega_{W}^{\rm halo}$, and the mean thermal energy density of baryons, $\Omega_{\rm th}$. These quantities can be obtained using two fundamentally different estimates: we compute $\Omega_{W}^{\rm halo}$ using the theoretical framework of the halo model which is driven by dark matter statistics, and measure $\Omega_{\rm th}$ using the Sunyaev-Zeldovich (SZ) effect which probes the mean thermal pressure of baryons. First, we derive that, at the present time, about 90\%  of $\Omega_{W}^{\rm halo}$ originates from massive halos with $M>10^{13}\,M_\odot$. Then, using our measurements of the SZ background, we find that $\Omega_{\rm th}$ accounts for about 80\% of the kinetic energy of the baryons available for pressure in halos at $z\lesssim 0.5$. This constrains the amount of non-thermal pressure, e.g., due to bulk and turbulent gas motion sourced by mass accretion, to be about $\Omega_{\rm non-th}\simeq 0.4\times 10^{-8}$ at $z=0$.
\end{abstract}


\section{Introduction} \label{sec:intro}

Inspired by the ``cosmic energy inventory'' program \citep{fukugita/peebles:2004}, we estimate the cosmic mean densities of two related forms of energies: the thermal ($\Omega_{\rm th}$) and gravitational potential energy ($\Omega_{W}$) in large-scale structure of the Universe. As structure formation proceeds, the gravitational energy associated with matter density fluctuations is converted into kinetic energy, most of which thermalizes via shocks \citep{cen/ostriker:1999,miniati/etal:2000}. Thus, the comparison between $\Omega_{\rm th}$ and $\Omega_{W}$ characterizes the efficiency of thermalization in large-scale structure. These quantities, which can be inferred in two fundamentally different ways,  provide an important consistency check of the cosmic energy inventory. In addition, precise enough estimates might allow us to constrain the amount of non-thermal energy, disentangle gravitational and non-gravitational heating, and assess the level of virialization of dark matter halos. 
This is the goal of our analysis.

Early estimates of the amount of thermal energy were introduced in \cite{cen/ostriker:1999,refregier/etal:2000,zhang/pen:2001,zhang/pen/trac:2004} using analytical models and cosmological hydrodynamical simulations.
Estimates of the amount of gravitational potential energy were presented by  \cite{fukugita/peebles:2004} which led to the first estimates of $\Omega_{W}$ for dark matter halos but for only two coarse halo mass bins.

It is now timely to revisit these estimates. First, our quantitative understanding of the large-scale structure of the Universe has significantly improved, in terms of the cosmological parameter estimation and, mainly, in terms of the dark matter halo statistics. Second, we now have access to observational constraints on $\Omega_{\rm th}$ as a function of redshift, using recent measurements of halo bias-weighted electron pressure of the Universe from the thermal Sunyaev-Zeldovichc (SZ) effect \citep{sunyaev/zeldovich:1972}. This quantity can be constrained by cross-correlating data from the Planck satellite \citep{Planck_2016_HFI,planck_sz:2016} with galaxies tracing the matter density field \citep{vikram/lidz/jain:2017,pandey/etal:2019,koukoufilippas/etal:2020}.
In particular, our recent work \citep[hereafter Paper~I]{chiang/etal:2020} 
probes this signal over a wider redshift range with better controlling systematic effects due to contamination of the Galactic foregrounds and the cosmic infrared background (CIB) than previous work. We compare these improved estimates of $\Omega_{\rm th}$ and $\Omega_{W}$ as a function of the cosmic time, and use them to infer the corresponding amount of non-thermal energy.

In this paper we focus on energy densities associated with the formation of large-scale structure, in the sense that we do not consider dissipative gravitational settling involved in star formation and accretion disks.
The rest of the paper is organized as follows. In Section \ref{sec:thermal}, we review our measurement of $\Omega_{\rm th}$ obtained in Paper I. In Section \ref{sec:grav}, we present our calculations of $\Omega_{W}$ from all large-scale structure of the Universe as well as that from collapsed structures (halos). In Section \ref{sec:LI}, we solve the Layzer-Irvine equation to find the relationship between the kinetic energy density and $\Omega_{W}$. We interpret the results and conclude in Section \ref{sec:conclusion}. 

Throughout, we use the Planck 2018 ``TT,TE,EE+lowE+lensing'' cosmological parameters given in Table 1 of \citet{planck2018_cosmo}: ($h$, $\Omega_{\rm c}h^2$, $\Omega_{\rm b}h^2$, $A_{\rm s}$, $n_{\rm s}$) = (0.6737, 0.1198, 0.02233, $2.097\times10^{-9}$, 0.9652), with the minimal mass for neutrinos (0.06~eV). The density parameter of the massive neutrinos is $\Omega_\nu h^2=0.06/93.14=6.4\times 10^{-4}$. Thus, the total mass density parameter is $\Omega_{\rm m}=\Omega_{\rm c}+\Omega_{\rm b}+\Omega_{\nu}=0.3146$ and the cosmological constant is given by $\Omega_\Lambda = 1 - \Omega_{\rm m}$. The mean baryon fraction in halos is given by $f_{\rm b}=\Omega_{\rm b}/(\Omega_{\rm c}+\Omega_{\rm b})=0.157$.

\section{Thermal energy density}\label{sec:thermal}

The density parameter for the comoving thermal energy density $\rho_{\rm th}$, which is related to the mean physical thermal pressure $\langle P_{\rm th}\rangle$, can be defined as\footnote{Note that we define the thermal energy density in terms of the thermal pressure. If we defined it in terms of the thermal kinetic energy instead, we should multiply  Eq.~(\ref{eq:thermaldef}) by 3/2.}
\begin{equation}
    \Omega_{\rm th}(z)\equiv \frac{\rho_{\rm th}(z)}{\rho_{\rm crit}} =  \frac{\langle P_{\rm th}(z) \rangle}{\rho_{\rm crit}\,(1+z)^3}\,, 
    \label{eq:thermaldef}
\end{equation}
where $\rho_{\rm crit}=1.054\times 10^4~h^2~{\rm eV~cm^{-3}}$ is the present-day critical energy density.

Following \cite{cen/ostriker:1999} and \cite{refregier/etal:2000}, an estimate of this quantity can be obtained by writing the pressure in terms of the gas density-weighted temperature defined by $k_{\rm B}\bar T_\rho\equiv k_{\rm B}\langle\rho_{\rm gas}T_{\rm gas}\rangle/\langle\rho_{\rm gas}\rangle=\langle P_{\rm gas}\rangle/\langle n_{\rm gas}\rangle$ where $k_{\rm B}$ is the Boltzmann constant, $n_{\rm gas}=(8-5\,Y)\rho_{\rm b}/(4\,m_{\rm p})$ with $m_{\rm p}$ being the proton mass and $Y$ the primordial Helium mass fraction. In this way, one obtains \citep[see Eq.~(4) of][]{zhang/pen/trac:2004}
\begin{equation}
\label{eq:Trho}
    \Omega_{\rm th}(z) = 1.78\times 10^{-8}~\frac{k_{\rm B}\bar T_\rho(z)}{0.2~{\rm keV}}\frac{\Omega_{\rm b}}{0.049}\,.
\end{equation}
These authors attempted to estimate $\bar T_\rho$ using hydrodynamical simulations.

In Paper~I we provided a detailed description of $\Omega_{\rm th}$. Assuming that the gas is fully ionized, which is valid for hot gas in galaxy groups and clusters, we can express the thermal gas pressure $P_{\rm th}$ in terms of that of the electrons by $P_{\rm th} = {(8-5Y)}/{(4-2Y)}\,P_{\rm e}$. For $Y=0.24$ \citep[][the uncertainty ($\pm 0.008$) is negligible for our analysis]{planck2018_cosmo}, $P_{\rm th} = 1.932\,P_{\rm e}$. The mean electron pressure $\langle P_{\rm e} \rangle$ is the quantity we constrained in Paper I. By measuring the large-scale correlation between tracers of matter overdensities (galaxies and quasars) and the SZ effect, one can constrain the halo bias-weighted electron pressure, $\langle bP_{\rm e}\rangle(z)$ \citep{vikram/lidz/jain:2017}, which is directly proportional to the observable $b_y \times dy/dz$:
\begin{equation}
\langle b P_{\rm e}\rangle = \langle P_{\rm e}\rangle b_y  = \frac{m_{\rm e}\, c^2\, (1+z)}{\sigma_{\rm T}}\, \frac{{\rm d}z}{{\rm d}\chi}\, \left (b_y\,\frac{{\rm d}y}{{\rm d}z} \right)\,,
\label{eq:bPe}
\end{equation}
where $\sigma_{\rm T}$ is the Thomson scattering cross section, $m_{\rm e}$ the electron mass, $c$ the speed of light, $\chi$ the comoving radial distance, $y$ the Compton $y$ parameter, and $b_y$ the SZ-weighted halo bias (see below). Inferring $\Omega_{\rm th}(z)$ thus requires estimating the two key quantities $\langle b P_{\rm e}\rangle$ and $b_y$. We present them below.

\subsection{Observational constraints from SZ measurements}

In Paper I, we estimated $\langle bP_{\rm e}\rangle(z)$ by measuring angular correlations between the SZ signal and galaxies and quasars tracing the matter density field. To do so, we measured angular two-point cross-correlation functions, $w(\theta,z)$, of intensity maps in microwave bands from the Planck mission \citep[at 100,  143,  217,  353,  545 and 857 GHz;][]{Planck_2016_HFI} and the locations of two million spectroscopic galaxies and quasars from the Sloan Digital Sky Survey \citep[SDSS;][]{2000AJ....120.1579Y} over $0<z\lesssim 3$. The spectroscopic redshift references include the main, the Baryon Oscillation Spectroscopic Survey, and the Extended Baryon Oscillation Spectroscopic Survey samples \citep{2001AJ....122.2267E,2002AJ....124.1810S,2005AJ....129.2562B,2010AJ....139.2360S,2016MNRAS.455.1553R,2017A&A...597A..79P,2018MNRAS.473.4773A,2018ApJ...863..110B}. We also use the reprocessed Infrared Astronomical Satellite (IRAS) data \citep{2005ApJS..157..302M} at 3 and 5~THz for a better separation of the CIB and the SZ effect. 

To probe the regime relevant to the large-scale structure formation, we focus on the large-scale limit of the two-point correlation function called the ``2-halo term'' \citep{cooray/sheth:2002}. We measured $w(\theta,z)$ integrated over $\theta_{\rm min}(z)<\theta<\theta_{\rm max}(z)$. The maximum angular scale is chosen as $\theta_{\rm max}(z)=8~{\rm Mpc}/D_{\rm A}(z)$ where $D_{\rm A}(z)$ is the proper angular diameter distance, so that the results are not affected by the systematic large-scale zero point fluctuations. The minimum angular scale is chosen as $\theta_{\rm min}(z)=3~{\rm Mpc}/D_{\rm A}(z)$ at 100 and 143~GHz and $2~{\rm Mpc}/D_{\rm A}(z)$ at higher frequencies, so that the measured cross-correlation is dominated by the 2-halo term \citep[see Figure 3 of][also see Appendix A of Paper I where we tested the combined effect of this plus non-linear galaxy bias]{vikram/lidz/jain:2017}. The larger $\theta_{\rm min}$ for 100 and 143~GHz is due to their larger beam sizes. We determined the best-fitting large-scale amplitudes of the SZ effect by marginalizing over the CIB parameters. 

Alternatively, one can use the small-scale limit of the two-point correlation function called the ``1-halo term'' (via, e.g., stacking analysis equivalent of the cross-correlation function on small angular separations), which is sensitive to the average thermal pressure profile of halos. This measurement can probe the redistribution of thermal energy in halos by baryonic feedback of, e.g., active galactic nuclei \citep[AGN;][]{planck_sz_stacking:2013,vanwaerbeke/etal:2014,greco/etal:2015,ruan/mcquinn/anderson:2015,ma/etal:2015,hojjati/etal:2015,hojjati/etal:2017,crichton/etal:2016,spacek/etal:2016,spacek/etal:2017,soergel/etal:2017,lim/etal:2018a,lim/etal:2018b,hall/etal:2019,tanimura/etal:2020}, though the contribution of the 2-halo term should still be taken into account when interpreting the measurement \citep{hill/spergel:2014,battaglia/etal:2015,hill/etal:2018}.

\subsection{Thermal pressure bias normalization}

\begin{figure*}[t]
 \begin{center}
 \includegraphics[width=12cm]{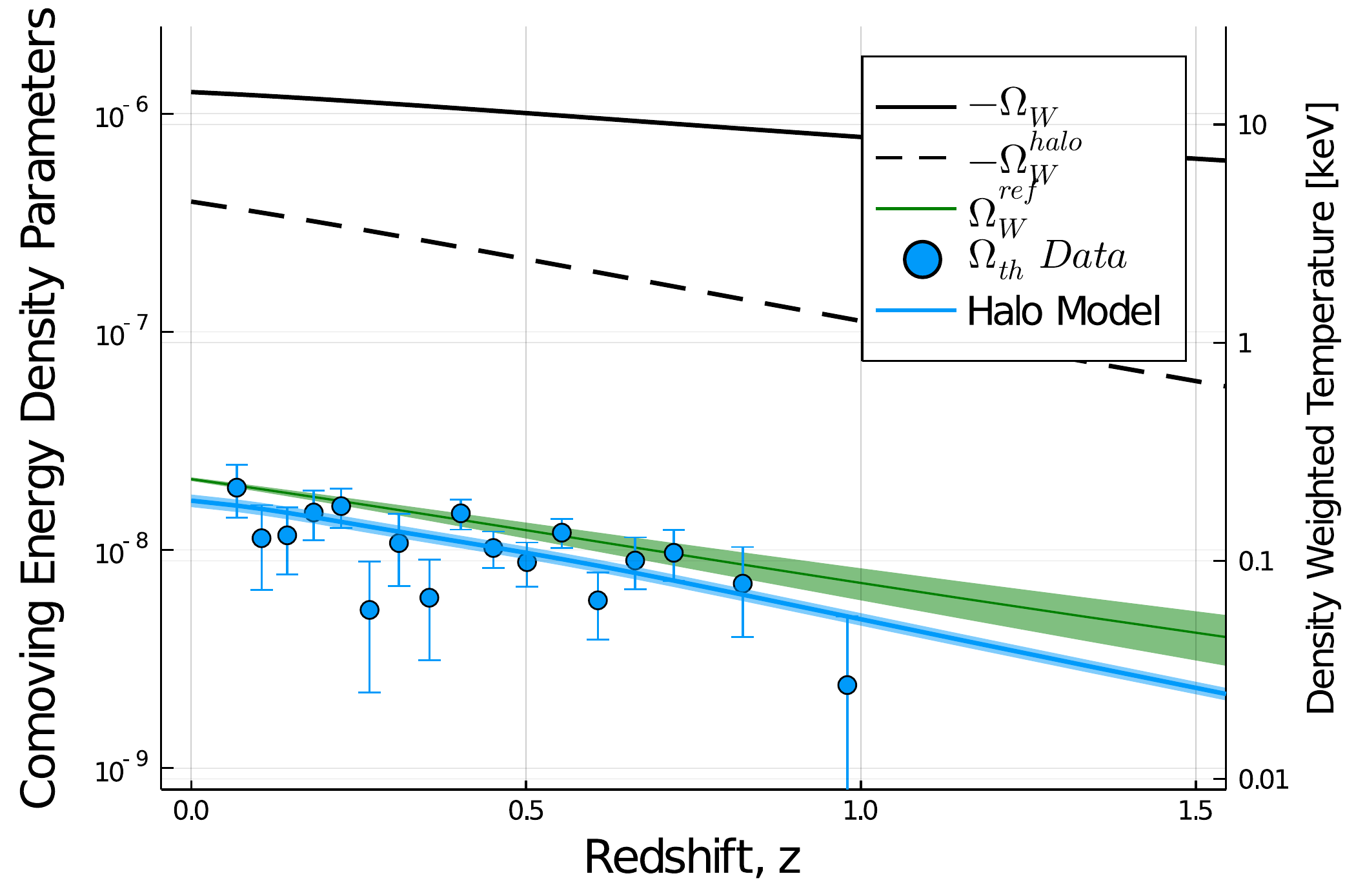}
\caption{
Comoving density parameters of the thermal energy (data points with error bars from Paper I), gravitational potential energy of all large-scale structure of the Universe (black solid line), and that of halos (black dashed line). The lower boundary of the green shaded area shows the halo contribution times the mean baryon fraction $f_{\rm b}=0.157$, whereas the upper boundary shows the difference between the corresponding non-linear and linear $P(k)$ contributions. The blue shaded area shows the best-fitting model for $\Omega_{\rm th}$ (Eqs.~(\ref{eq:thermaldef}, \ref{eq:bPdef}, \ref{eq:ET}, \ref{eq:by})) with a constant mass bias parameter and its 68\% confidence level, $B=1.27_{-0.04}^{+0.05}$. The right axis shows the corresponding values of $\bar T_{\rho}$ for $\Omega_{\rm th}$ given in Eq.~(\ref{eq:Trho}).}
\label{fig:omega_grav}
\end{center}
\end{figure*}

We now explain our halo model, which allows us to (i) calculate and correct for $b_y$ from the observed 2-halo term angular correlation to derive the mean pressure $\langle P_{\rm e}\rangle$, and (ii) interpret the amplitude of $\langle P_{\rm e}\rangle$ as contributions from halos with the halo mass-pressure normalization quantified by a mass bias parameter.

Following the steps presented in Paper~I, we use a model for the 2-halo term \citep[correlation between two halos, see][]{komatsu/kitayama:1999} of the angular cross power spectrum. The angular cross-correlation function of the SZ and spectroscopic galaxy data at a given redshift, $w(\theta,z)$,  is related to the angular cross power spectrum, $C_\ell(z)$, as $w(\theta,z) = (2\pi)^{-1}\int_0^\infty \ell {\rm d}\ell~C_\ell(z) J_0[(\ell+1/2)\theta]$, 
where $J_0(x)$ is the Bessel function of order 0. We model the 2-halo term as \citep[e.g.,][]{makiya/ando/komatsu:2018}
\begin{equation}
    C_\ell^{\rm 2h}(z_i) = 
    \int {\rm d}z'\frac{{\rm d}V}{{\rm d}z' {\rm d}\Omega}f_\ell^{y}(z')f_\ell^g(z',z_i)P(k_\ell,z')\,,
\end{equation}
where ${\rm d}V/{\rm d}z{\rm d}\Omega$ is the comoving volume element per unit solid angle, $k_\ell\equiv (\ell+1/2)/[(1+z)D_{\rm A}(z)]$, $P(k,z)$ is the matter power spectrum, and $z_i$ is the mean redshift of the $i$th bin.
The functions $f_\ell^{g}$ and $f_\ell^{y}$ are given in Eq.~(23) of \citet{makiya/ando/komatsu:2018} and
Eq.~(5) of \citet{komatsu/kitayama:1999}, respectively.

On scales larger than the size of halos, both $f_\ell^g$ and $f_\ell^y$ approach $\ell$-independent values. The former becomes $f_0^g(z,z_i)=b_g(z)W^g(z,z_i)/\chi^2(z)$, where $b_g(z)$ is a linear galaxy bias parameter, $W^g(z,z_i)$ is the normalized redshift distribution of galaxies within the $i$th bin, and $\chi(z)=(1+z)D_{\rm A}(z)$ is the comoving distance out to redshift $z$. We constrain $f_0^g$ from the auto galaxy power spectra of SDSS.
The latter quantity $f_\ell^y$ yields a halo bias-weighted electron pressure, $\langle bP_{\rm e}\rangle$, as 
\begin{equation}
f_0^y(z) = \frac{\sigma_{\rm T}}{m_{\rm e}c^2}{\frac{\langle bP_{\rm e}\rangle(z)}{(1+z)\chi^2(z)}}\,,
\end{equation}
where 
\begin{eqnarray}
\label{eq:bPdef}
\langle bP_{\rm e}\rangle(z)&\equiv& (1+z)^3\int_{M_{\rm min}}^{M_{\rm max}} {\rm d}M\frac{{\rm d}n}{{\rm d}M}b_{\rm lin}E_{\rm T}\,,
\end{eqnarray}
with
\begin{eqnarray}
\label{eq:ET}
E_{\rm T}(M,z)&\equiv& 4\pi\int_0^{r_{\rm max}} r^2{\rm d}rP_{\rm e}(r,M,z)\,
\end{eqnarray}
being the thermal energy of electrons per halo as defined by \cite{vikram/lidz/jain:2017}. Here, ${\rm d}n/{\rm d}M$ and $b_{\rm lin}$ are the comoving number density \citep{tinker/etal:2008} and linear bias  \citep{tinker/etal:2010} of halos, respectively.
The halo bias-weighted electron pressure 
is a direct observable of the SZ-galaxy cross correlation, which requires no assumption about ${\rm d}n/{\rm d}M$ or $b_{\rm lin}$.

To calculate $b_y$, we follow methodology given in \citet{makiya/ando/komatsu:2018,makiya/hikage/komatsu:2020}.\footnote{Codes are available in \url{https://github.com/ryumakiya/pysz}.} Its expression is given by
\begin{equation}
b_y(z) =
\frac{\int {\rm d}M ({\rm d}n/{\rm d}M) \tilde{y}_{0}(M,z)b_{\rm lin}(M,z)}
{\int {\rm d}M ({\rm d}n/{\rm d}M) \tilde{y}_{0}(M,z)}\,,
\label{eq:by}
\end{equation}
with $\tilde{y}_0=\sigma_{\rm T}E_{\rm T}/(m_{\rm e}c^2D_{\rm A}^2)$. 
Here, $\tilde{y}_\ell$ is the Fourier transform of the 2D profile of the SZ effect within a halo given in Eq.~(2) of \citet{komatsu/seljak:2002}.
To calculate $\tilde{y}_0$ we use the universal pressure profile derived from the X-ray and Planck data, which gives $\tilde{y}_0\propto M^{5/3+\alpha_{\rm p}}$ with $\alpha_{\rm p}=0.12$ \citep{arnaud/etal:2010,planck_inter_v:2013}. We may write this as $\tilde{y}_0\propto GM^{2+\alpha_{\rm p}}/R$ where $R\propto M^{1/3}$; thus, Eq.~(\ref{eq:by}) can also be seen as gravitational potential energy-weighted halo bias.

The mass $M$ in Eq.~(\ref{eq:by}) is $M_{500}$, for which the mean overdensity within a halo is 500 times the critical density of the Universe. The ${\rm d}n/{\rm d}M$ of \citet{tinker/etal:2008} is given for halo masses defined using various overdensities with respect to the mean mass density of the Universe, $200\le \Delta_{\rm m}\le 3200$. We thus interpolate the coefficients of ${\rm d}n/{\rm d}M$ for an overdensity of $\Delta_{\rm m}=500\, E^2(z)/[\Omega_{\rm m}(1+z)^3]$, where $E^2(z)\equiv\Omega_{\rm m}\,(1+z)^3+\Omega_\Lambda$. We then integrate Eq.~(\ref{eq:by}) over $M_{500}$ from $M_{\rm min}=10^{11}~M_\sun/h$ to $M_{\rm max}=5\times 10^{15}~M_\sun/h$. Lowering $M_{\rm min}$ to $10^{10}~M_\sun/h$ changes $\rho_{\rm th}$ by less than 1, 3, and 10 percent at $z<1$, 2, and 3, respectively. These changes are smaller than the theoretical uncertainty of ${\rm d}n/{\rm d}M$ in the corresponding redshift ranges.

To capture all the thermal pressure associated with a halo, we integrate the pressure profile out to $r_{\rm max}=6\,r_{500}$, where $r_{500}$ is the radius that encloses $M_{500}=(4\pi/3)\,E^2(z)\,\rho_{\rm crit}\,r_{500}^3$. This radius approximates the so-called shock radius, beyond which pressure falls rapidly \citep{bryan/norman:1998,lau/etal:2015,shi:2016,zhang/etal:2020}. For $r_{\rm max}=4\,r_{500}$ and $8\,r_{500}$, we find 0.929 and 1.036 times $\rho_{\rm th}$ for $r_{\rm max}=6\,r_{500}$, which change the inferred values of the mass bias parameter $B$, which affects the normalization of the mass-pressure relationship (see below), by 4 and 2 percent, respectively.

Combining our measurements of $\langle b P_{\rm e}\rangle$ and the estimation of $b_y$, we obtain estimates of $\Omega_{\rm th}(z)$, shown with the data points in Figure~\ref{fig:omega_grav}. In particular, we find
\begin{equation}
\Omega_{\rm th} = (1.7\pm0.1) \times 10^{-8}~~{\rm at~}z=0\;,
\end{equation}
from the best-fitting model with the 68\% confidence level. The best estimate and uncertainty here were derived assuming a constant mass bias parameter $B$ using the posterior reported in Paper~I. If we allow for a redshift-dependent $B$, we find $\Omega_{\rm th} = (1.5\pm0.3) \times 10^{-8}$ at $z=0$. As there is no evidence for the evolution of $B$ within the redshift range discussed here, we use the redshift-independent value as the baseline in this paper. 
Since we measure $\langle b P_{\rm e}\rangle$ directly over $0\lesssim z \lesssim 1$, and the estimate of $b_y$ does not depend on the mass bias $B$, our $\Omega_{\rm th}$ estimates within this redshift range should be considered empirical.
We have checked the accuracy of the analytical model for $b_y$ using the Magneticum simulation \citep{dolag/komatsu/sunyaev:2016} and found that the uncertainty is negligible compared with the statistical uncertainty of $\Omega_{\rm th}$.

To interpret the observed $\Omega_{\rm th}$ as the sum of halo contributions with the mass-pressure calibration not precisely known, we parameterize this uncertainty using the so-called ``mass bias parameter'' $B$. It relates the true $M_{500}$ of halos to the mass calibrated by X-ray observations \citep{arnaud/etal:2010} used by the Planck team \citep{planck_inter_v:2013} as $B=M^{\rm true}_{500}/M^{\rm Planck}_{500}$. It is found that $B>1$ is required to fit statistics of the Planck SZ data such as number counts and auto- and cross-power spectra \citep{planck_sz_cosmo:2014,planck_sz_cosmo:2016,dolag/komatsu/sunyaev:2016,hurier/lacasa:2017,salvati/etal:2018,salvati/etal:2019,bolliet/etal:2018,bolliet/etal:2020,makiya/ando/komatsu:2018,makiya/hikage/komatsu:2020,osato/etal:2018,osato/etal:2020,koukoufilippas/etal:2020}. This parameter is related to the commonly used parameter $b$ as $1-b=B^{-1}$ \citep[see also][for yet another parameterization]{horowitz/seljak:2017}. Assuming a mass- and redshift-independent $B$, which is supported by the current measurements \citep{makiya/ando/komatsu:2018,salvati/etal:2019,koukoufilippas/etal:2020,chiang/etal:2020}, we find $B=1.27_{-0.04}^{+0.05}$ from our measurement of $\langle bP_{\rm e}\rangle$ (Paper I). The $B$ dependence is simple: $\Omega_{\rm th}\propto \langle bP_{\rm e}\rangle\propto B^{-5/3-\alpha_{\rm p}}$ with $\alpha_{\rm p}=0.12$, as the total electron pressure is proportional to $M^{5/3+\alpha_{\rm p}}$ \citep{arnaud/etal:2010} with $M=M_{500}^{\rm Planck}=M_{500}^{\rm true}/B$.

The physical origin of $B$ is not fully understood. The origin of $B>1$ is partly due to the presence of non-thermal gas pressure in galaxy clusters, which leads to underestimation of X-ray calibrated mass assuming hydrostatic equilibrium between gravity and thermal gas pressure gradient \citep{kay/etal:2004,rasia/etal:2006,rasia/etal:2012,nagai/vikhlinin/kravtsov:2007,jeltema/etal:2008,piffaretti/valdarnini:2008,lau/kravtsov/nagai:2009,meneghetti/etal:2010,nelson/etal:2012,nelson/etal:2014,shi/komatsu:2014,shi/etal:2016,biffi/etal:2016,henson/etal:2017,vazza/etal:2018,angelinelli/etal:2020,ansarifard/etal:2020}. It is possible that some of $B>1$ is due to the choice of particular X-ray cluster samples and analysis methods used in \citet{arnaud/etal:2010} and/or \citet{planck_inter_v:2013}. We therefore choose to call this parameter simply ``mass bias'', without reference to the assumption of hydrostatic equilibrium.

\section{Gravitational potential energy density}\label{sec:grav}

We now estimate the density parameter for the gravitational potential energy as
\begin{equation}
\label{eq:omegaWdef}
    \Omega_{W} = \frac{\Omega_{\rm m}}{c^2}\,W\;,
\end{equation}
where $W$ is the gravitational potential energy per unit mass. Considering a system of mass $M$ consisting of particles with mass $m_i$, such that $M=\sum_i m_i$, we can express its gravitational potential energy as
\citep[Section 24 of][]{peebles:1980}
\begin{eqnarray}
\label{eq:MW}
 MW &=& -\frac12a^3\rho_{\rm m}(a)\int d^3x~\delta({\bf x},a)\phi({\bf x},a)\\
 \nonumber
&=& -\frac12Ga^5\rho_{\rm m}^2(a)\int d^3x\int d^3x'~\frac{\delta({\bf x},a)\delta({\bf x}',a)}{|{\bf x}-{\bf x}'|}\,,
\end{eqnarray}
where $a(t)=(1+z)^{-1}$ is the scale factor, ${\bf x}$ denotes the comoving coordinates, $\delta({\bf x},a)$ is the matter density contrast, and $\phi({\bf x},a)$ is the peculiar gravitational potential. The factor $1/2$ corrects double-counting of particles contributing to both $\delta$ and $\phi$.

For a uniform density sphere with a physical radius $R_{\rm p}$, an excess mass above the mean $\delta M=(4\pi/3)\delta\rho_{\rm m} R_{\rm p}^3$, and a potential $\phi(r_{\rm p}\le R_{\rm p})=-2\pi G\delta\rho_{\rm m}(R_{\rm p}^2-r_{\rm p}^2/3)$, we find $MW = -3G(\delta M)^2/(5R_{\rm p})$, which agrees with the known result for gravitational potential energy of a uniform density sphere. 

Going to Fourier space and taking ensemble average of Eq.~(\ref{eq:MW}), we obtain
\begin{equation}
\label{eq:MWfourier}
 MW = -\frac12\rho_{\rm m0}\left(\int d^3x\right)
\int \frac{d^3k}{(2\pi)^3}P_{\phi\delta}(k,a)\,,
\end{equation}
where $\rho_{\rm m0}=\rho_{\rm m}a^3$ is the present-day mean mass density. The cross-correlation power spectrum of $\phi$ and $\delta$,
$P_{\phi\delta}(k)$, is related to the matter density power spectrum $P(k)$ via the Poisson equation:
\begin{equation}
 P_{\phi\delta}(k,a) = -4\pi G\frac{\rho_{\rm m0}}{a} \frac{P(k,a)}{k^2}\,.
\end{equation}
Dividing both sides of Eq.~(\ref{eq:MWfourier}) by $M=\rho_{\rm m0}\int d^3x$, we obtain the gravitational potential energy per unit mass of large-scale structure of the Universe. Using the Friedmann equation, $H_0^2=(8\pi G/3)\rho_{\rm crit}$, and $\rho_{\rm m0}=\Omega_{\rm m}\rho_{\rm crit}$, we find
\begin{eqnarray}
\label{eq:W}
W  &=& -\frac{3\Omega_{\rm m}H_0^2}{8\pi^2 a}\int_0^\infty {\rm d}k~P(k,a)\,.
\end{eqnarray}
Multiplying this by $\Omega_{\rm m}/2$ agrees with Eq.~(60) of \citet{fukugita/peebles:2004} for the binding energy for the present-day epoch with $a=1$. 

The gravitational potential energy $W$ per unit mass, or equivalently, the density parameter $\Omega_W$, is \emph{exactly} given by the integral over the matter power spectrum $P(k)$. It does not rely on any assumption and is valid irrespective of other aspects of the density field, such as non-Gaussian or higher-order statistical properties.

When we calculate $W$ from all forms of matter in the Universe, we use $\rho_{\rm m0}$ and $P(k)$ including baryons, cold dark matter (CDM), and neutrinos. When we calculate $W$ from halos in the next section, we exclude the contribution from neutrinos as they cluster on halo scales only weakly. This simple prescription has been shown to work well for the halo mass function \citep{ichiki/takada:2012,costanzi/etal:2013}, bias and power spectrum \citep{villaescusa-Navarro/etal:2014,castorina/etal:2014}, and statistics of the SZ effect \citep{bolliet/etal:2020}.
Thus, for halo-based calculations, we replace $\rho_{\rm m0}$ with $\rho_{\rm cb0}\equiv \rho_{\rm b0}+\rho_{\rm c0}$, and use the baryon+CDM power spectrum, $P_{\rm cb}(k)$, to compute all the ncessary ingredients of the halo model.

\subsection{Contribution from halos}

The total gravitational potential energy density in the Universe includes contributions from all large-scale structures, which can be evaluated using the non-linear matter $P(k)$ in Eq.~(\ref{eq:W}). In contrast, as shown in Paper I, the total thermal energy density in the Universe is dominated by collapsed halos, especially massive clusters and groups. To meaningfully compare the two, we need to estimate the contribution of $\Omega_W$ originating from halos. 

To this end, we use the halo model \citep{cooray/sheth:2002} to write
$P(k) = P_{\rm 1h}(k)+P_{\rm 2h}(k)$, 
where $P_{\rm 1h}$ and $P_{\rm 2h}$ are the 1- and 2-halo terms, respectively. In this section we focus on the 1-halo term, to extract $\Omega_W$ originating from halos. The 1-halo term is given by \citep{seljak:2000}
\begin{equation}
\label{eq:P1h}
 P_{\rm 1h}(k,a) = \frac1{\rho_{\rm cb0}^2}\int_{M_{\rm min}}^{M_{\rm max}} {\rm d}M~\frac{{\rm d}n}{{\rm d}M}M^2|u(kR,c)|^2\,,
\end{equation}
where $u(x,c)$ is the Fourier transform of the halo mass density profile to be specified, normalized such that $u(x)\to 1$ for $x\to 0$, and $c$ is the concentration parameter\footnote{Here, $u(x,c)$ is the total mass density profile and we do not yet distinguish between the density profiles of dark matter and baryons. Later we adopt a Navarro-Frenk-White (NFW) profile \citep{navarro/frenk/white:1996,navarro/frenk/white:1997} for $u(x,c)$, which is a good approximation to the dark matter density profile. On the other hand the baryonic matter distribution is less concentrated than the dark matter \citep{schaan/etal:prep}. We comment on this uncertainty in Section~\ref{sec:accuracy}.}.
Here, $R$ is the characteristic comoving radius of a halo, e.g., the comoving virial radius. It is related to the mass enclosed within $R$ as $M=(4\pi/3)\rho_{\rm cb0}\Delta_{\rm m} R^3$, where $\Delta_{\rm m}$ is the overdensity of a halo with respect to the mean mass density of the Universe. We use a comoving radius rather than a physical one, as $k$ in the power spectrum is the comoving wavenumber.

Using $P_{\rm 1h}(k)$ in Eq.~(\ref{eq:W}), we obtain
\begin{equation}
 W_{\rm 1h}  = -\frac1{\pi a}
\frac1{\rho_{\rm cb0}}\int {\rm d}M~\frac{{\rm d}n}{{\rm d}M}\frac{GM^2}{R}\int_0^\infty {\rm d}x~|u(x,c)|^2\,.
\end{equation}
We may use the physical radius, $R_{\rm p}=aR$, to re-write this result as
\begin{equation}
\label{eq:physics}
W_{\rm 1h}  = -\frac1{\rho_{\rm cb0}}\int_{M_{\rm min}}^{M_{\rm max}} {\rm d}M~\frac{{\rm d}n}{{\rm d}M}\frac{GM^2}{R_{\rm p}}A_g(c)\,,
\end{equation}
where 
\begin{equation}
\label{eq:Ag}
 A_g(c)\equiv \int_0^\infty \frac{{\rm d}x}{\pi}~|u(x,c)|^2\,.
\end{equation}

\begin{figure}[t]
 \includegraphics[width=8.5cm]{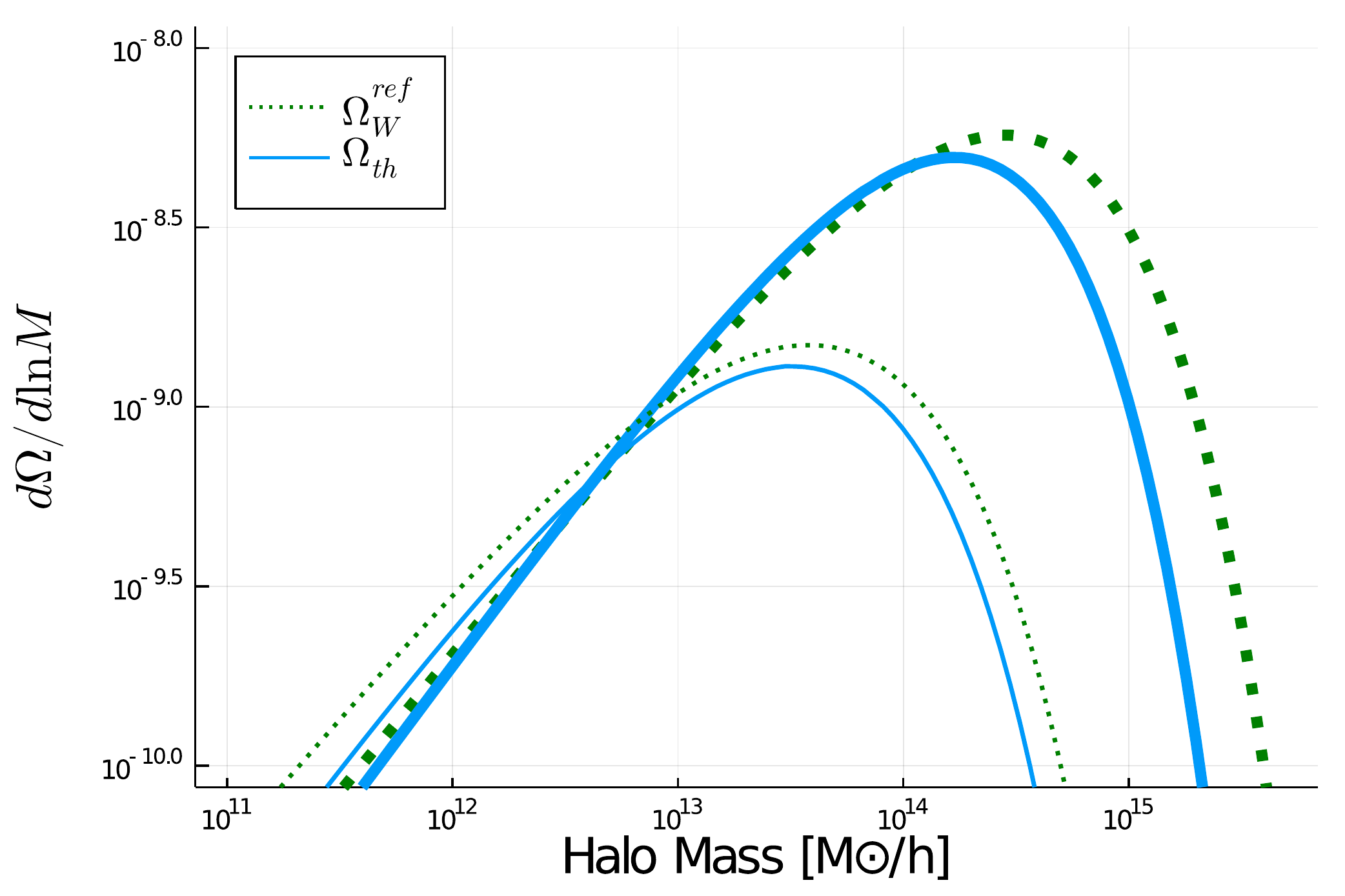}
\caption{
Contribution to $\Omega_{W}^{\rm ref}=-f_{\rm b}\Omega_{W}^{\rm halo}/3$ (Eq.~(\ref{eq:omega_grav_ref}); dotted lines) and $\Omega_{\rm th}$ (solid lines) per logarithmic mass interval, at $z=0$ (thick lines) and $1$ (thin lines). The mass for $\Omega_{W}^{\rm halo}$ is computed with $\Delta_{\rm m}=200$, while $\Omega_{\rm th}$ is with $M_{500}$.
\label{fig:dodlnM}
}
\end{figure}

Eq.~(\ref{eq:physics}) has a clear physical meaning. The Fourier transform of the density profile of a uniform density sphere is $u(x) = 3j_1(x)/x$, where $j_1(x)=[\sin(x)-x\cos(x)]/x^2$ is the spherical Bessel function of order 1 and $x=kR$. Integrating over $x$, we obtain $A_g=3/5$. Thus, $-A_gGM^2/R_{\rm p}=-3GM^2/(5R_{\rm p})$ agrees with the gravitational potential energy of a uniform density sphere with a physical radius $R_{\rm p}$.
Therefore, $W_{\rm 1h}$ is the mean gravitational potential energy density of halos per unit mass. The contribution as a function of halo mass is shown in Figure~\ref{fig:dodlnM} (dotted lines) at $z=0$ and $1$.

Figure~\ref{fig:dodlnM} also shows that the thermal energy density has a similar mass scaling, as 
\begin{eqnarray}
    \Omega_{\rm th}
    \propto \int {\rm d}M\frac{{\rm d}n}{{\rm d}M}E_{\rm T}
    \propto  \int {\rm d}M\frac{{\rm d}n}{{\rm d}M}\frac{GM^{2+\alpha_{\rm p}}}{R_{\rm p}}\,,
\end{eqnarray}
where $E_{\rm T}$ is the thermal energy of electrons per halo defined in Eq.~(\ref{eq:ET}) and $\alpha_{\rm p}=0.12$. The contributions to $\Omega_{W}^{\rm halo}$ and $\Omega_{\rm th}$ per logarithmic mass interval differ only slightly due to the small departure from the self-similar behavior quantified by $\alpha_{\rm p}$ and the different mass definitions used ($M_{200\rm m}$ for $\Omega_{W}^{\rm halo}$ and $M_{500}$ for $\Omega_{\rm th}$).

\subsection{Numerical results}

To evaluate Eq.~(\ref{eq:W}), we use three power spectra:\footnote{Julia codes are available in \url{https://github.com/komatsu5147/OmegaGrav.jl}.}
(1) the linear matter power spectrum from the public Boltzmann solver {\sf CLASS} \citep{blas/lesgourgues/tram:2011}, (2) the non-linear matter power spectrum from ``Halofit'' \citep{smith/etal:2003} with the parameters updated by \citet{takahashi/etal:2012}, and (3) the 1-halo term, $P_{\rm 1h}(k)$.
For (1) and (2) we use $P(k)$ including baryons, CDM, and neutrinos. For (3) we exclude the contribution from neutrinos.

For ${\rm d}n/{\rm d}M$ we use \citet{tinker/etal:2008} with $\Delta_{\rm m}=200$. For the halo density profile, we use an  NFW profile \citep{navarro/frenk/white:1996,navarro/frenk/white:1997}. For this profile we use the analytical result for Fourier transform given in \citet{scoccimarro/etal:2001} and integrate Eq.~(\ref{eq:Ag}) to obtain $A_g(c)$. To obtain this analytical Fourier transform, they assume that the halo density profile is truncated at $R=c\,r_{\rm s}$ where $r_{\rm s}$ is a comving NFW scale radius. In our setup, $R=R_{\rm 200m}$ within which the mean overdensity of a halo is $\Delta_{\rm m}=200$ times the mean mass density of the Universe. 

This treatment adds a subtlety: what if we truncate the density profile at a larger radius? $W_{\rm 1h}$ would increase. At the same time, such halo outskirts may not be virialized completely; thus, it is not clear how to count the contribution in this region. For this reason we regard $W_{\rm 1h}$ computed here as a lower bound for the mean gravitational potential energy of halos per unit mass. An upper bound comes from the difference between the non-linear and linear $P(k)$ contributions, which can be computed without ambiguity about the halo boundary.

We find a convenient fitting formula for $A_g(c)\simeq 1.0202 + \sum_{i=1}^4 a_i(c-5)^i$ with $a_i=5.376\times 10^{-2}$, $-1.842\times 10^{-3}$, $9.489\times 10^{-5}$, and $-2.352\times 10^{-6}$ for $i=1$, 2, 3, and 4, respectively. This formula is accurate to better than 0.1 percent everywhere in $2\le c\le 20$. 

For $c(M,z)$ we use the formula given in \citet{duffy/etal:2008} for $\Delta_{\rm m}=200$:
$c(M,z) = 10.14~(M/2\times 10^{12}~M_\sun/h)^{-0.081}(1+z)^{-1.01}$. 
Thus, $A(c)$ is greater than 3/5, which means that an NFW halo is more tightly bound than a uniform density sphere. For the mass integral in Eq.~(\ref{eq:P1h}), we use the same $M_{\rm min}$ and $M_{\rm max}$ as in Section \ref{sec:thermal}.

We integrate $\int {\rm d}k\,P(k)$ from $k=5\times 10^{-4}$ to $30~h~{\rm Mpc}^{-1}$, in which the halofit model is calibrated against the simulation. Extending to $k_{\rm max}=100~h~{\rm Mpc}^{-1}$ increases the result for non-linear $P(k)$ only by 0.5 percent at $z=0$ and less at higher redshifts and for linear $P(k)$.

\begin{deluxetable*}{ccc|cccc|hcc}
\tablecaption{The density parameters of the gravitational potential energy (Eq.~(\ref{eq:omegaWdef})), $\Omega_{W}$, for the linear ($\Omega_{W}^{\rm lin}$), non-linear ($\Omega_{W}^{\rm nl}$), and 1-halo $P(k)$ ($\Omega_{W}^{\rm halo}$). The 1st column shows the redshift $z$, while the 2nd, 3rd and 4th columns show $-10^7\Omega_{W}^{\rm lin}/2$, $-10^7\Omega_{W}^{\rm nl}/2$ and $-10^7\Omega_{W}^{\rm halo}/2$, respectively.
The 5th, 6th and 7th columns show $-10^7\Omega_{W}^{\rm halo}/2$ from halos with the mass $M>10^{12}$, $10^{13}$ and $10^{14}~M_\sun/h$, while the 8th and 9th columns show $-10^8f_{\rm b}\Omega_{W}^{\rm halo}/3$ and $-10^8f_{\rm b}(\Omega_{W}^{\rm nl}-\Omega_{W}^{\rm halo}$)/3, respectively. See Section~\ref{sec:accuracy} for the modelling uncertainties in $\Omega_{W}^{\rm lin}$ (2 percent for all $z$), $\Omega_{W}^{\rm nl}$ (10 percent for all $z$), and $\Omega_{W}^{\rm halo}$ (7 to 14 percent from $z=0$ to 1.5). The redshift points are chosen to allow for the precise interpolation between them.
\label{tab:results}}
\tablehead{
\colhead{$z$} & \colhead{$-10^7\Omega_{W}^{\rm lin}/2$} & \colhead{$-10^7\Omega_{W}^{\rm nl}/2$} & \colhead{$-10^7\Omega_{W}^{\rm halo}/2$} 
&  \colhead{$>10^{12}$}
&  \colhead{$>10^{13}$}
&  \colhead{$>10^{14}~M_\sun/h$}
& \nocolhead{$-10^8f_{\rm b}\Omega_{W}^{\rm halo}$}
& \colhead{$-10^8f_{\rm b}\Omega_{W}^{\rm halo}/3$}
& \colhead{$-10^8f_{\rm b}(\Omega_{W}^{\rm nl}-\Omega_{W}^{\rm lin})/3$}
}
\startdata
0.0 & 4.23 & 6.28 & 1.98 & 1.96 & 1.84 & 1.26 & 3.11 & 2.07 & 2.15 \\
0.3 & 4.00 & 5.55 & 1.39 & 1.37 & 1.23 & 0.68 & 2.19 & 1.46 & 1.62 \\
0.5 & 3.76 & 5.03 & 1.08 & 1.05 & 0.91 & 0.41 & 1.69 & 1.13 & 1.34 \\
0.7 & 3.50 & 4.54 & 0.83 & 0.80 & 0.65 & 0.24 & 1.30 & 0.87 & 1.10 \\
1.0 & 3.12 & 3.91 & 0.56 & 0.53 & 0.39 & 0.089 & 0.88 & 0.59 & 0.82 \\
1.3 & 2.80 & 3.39 & 0.38 & 0.35 & 0.22 & 0.029 & 0.60 & 0.40 & 0.62 \\
1.5 & 2.61 & 3.11 & 0.30 & 0.26 & 0.15 & 0.013 & 0.47 & 0.31 & 0.52\\
\enddata
\end{deluxetable*}

In Figure~\ref{fig:omega_grav} we show our estimate of $\Omega_{W}$ for the non-linear $P(k)$ (solid line) and 1-halo $P(k)$ (dashed line). We find
\begin{equation}
    \frac{ \Omega_{W}^{\rm halo} } {\Omega_{W}^{\rm tot}} \simeq 0.3~~{\rm at~}z=0\;.
\end{equation}
This implies that, at the present time, there is substantially more potential energy in the large-scale structure (or 2-halo component) than that in gravitationally bound halos (the 1-halo component). The numerical values are given in Table~\ref{tab:results}. The 1-halo term's contribution is 30 percent that of the non-linear $P(k)$ at $z=0$, and declines to 15 percent at $z=1$. The 5, 6 and 7th columns show the halo contributions from $M>10^{12}$, $10^{13}$, and $10^{14}~M_\sun/h$. We find that most of the contributions to $\Omega_{W}^{\rm halo}$ come from massive halos: 65 and 90 percent from $M>10^{14}$ and $10^{13}~M_\sun/h$ at $z=0$, respectively, and 70 percent from $M>10^{13}~M_\sun/h$ at $z=1$. 

The 1-halo contribution integrated up to $R=R_{\rm 200m}$ agrees well with the difference between the non-linear and linear $P(k)$ contributions at $z=0$, while it is somewhat smaller at higher redshifts. The lower boundary of the green shaded area shows the 1-halo $P(k)$ times 2/3 the mean baryon fraction, whereas the upper boundary shows the difference between the non-linear and linear $P(k)$ contributions times 2/3 the mean baryon fraction. The green shaded area in Figure~\ref{fig:omega_grav} therefore shows the lower and upper bounds for a contribution in potentially non-virialized regions of halo outskirts.

\subsection{Accuracy and uncertainty}
\label{sec:accuracy}

\cite{fukugita/peebles:2004} previously provided an order-of-magnitude estimate of the gravitational potential energy in large-scale structure using the non-linear $P(k)$ estimate available at the time. 
Thanks to an improved understanding of non-linear matter clustering and precision determination of the cosmological parameters today, we can now estimate $\Omega_{W}$ with greater accuracy and the meaningful error estimates.

The dominant uncertainty for $\Omega_{W}$ from the linear $P(k)$ is the primordial scalar amplitude parameter $A_{\rm s}$, which is known to better than 2 percent precision \citep{planck2018_cosmo}. Uncertainty for $\Omega_{W}$ from the non-linear $P(k)$ is larger than this because accuracy of the halofit $P(k)$ at $k>1~h~{\rm Mpc}^{-1}$ is at the level of 5--10 percent depending on the wavenumber \citep{takahashi/etal:2012}. We thus quote 7 percent uncertainty for $\Omega_{W}$ from the non-linear $P(k)$ at $z\le 1.5$. The uncertainty at higher $z$ is smaller due to smaller non-linearity.

\citet{fukugita/peebles:2004} gave rough estimates of $\Omega^{\rm halo}_{W}$ for halos of typical $L^*$ galaxies and clusters of galaxies. Our calculation of the halo contribution to $\Omega^{\rm halo}_{W}$ based on the 1-halo $P(k)$ term includes contributions from halos of all masses.
The accuracy of this calculation is limited by the uncertainties of ${\rm d}n/{\rm d}M$. The precision of ${\rm d}n/{\rm d}M$ at $z=0$ is 5 percent \citep{tinker/etal:2008}. The uncertainty increases for larger redshifts because the time dependence of ${\rm d}n/{\rm d}M$ is still not understood very well. We thus quote 10 percent uncertainty for $\Omega^{\rm halo}_{W}$ at $z=1.5$. Nonetheless, we find that $\Omega_{W}$ computed with ${\rm d}n/{\rm d}M$ from \citet{tinker/etal:2008} and \citet{tinker/etal:2010} agree to within 2 percent at $z\le 1.5$, if the halo multiplicity function of \citet{tinker/etal:2010} is re-normalized to unity at each $z$.

Other sources of uncertainty include the effect of baryonic feedback on the matter power spectrum \citep{vanDaalen/etal:2011}, halo profile \citep{gredin/etal:2004}, and ${\rm d}n/{\rm d}M$ \citep{bocquet/etal:2016}. The uncertainty due to these effects is comparable to the current uncertainty for our calculations of $\Omega_{W}$ from the non-linear $P(k)$ and halos. We thus add the uncertainties in quadrature.

To summarize, the uncertainties in $\Omega_{W}$ from the linear and non-linear $P(k)$ are 2 and 10 percent, respectively, at all $z$, whereas that from the halos increases from 7 to 14 percent at $z=0$ to 1.5.

\section{Thermal and gravitational potential energies}
\label{sec:LI}

\subsection{Virial relation for halos}

One  can relate  the  kinetic energy of the gas to the gravitational potential energy, which is dominated by collisionless dark matter particles. For virialized structures such as halos, we can use the virial theorem,
\begin{equation}
\label{eq:virial}
    K = -\frac{W}{2}\;.
\end{equation}
where $K$ is the mean kinetic energy per unit mass, $K=\sum_im_iv_i^2/(2\sum_im_i)$. 

Without depending on the validity of the virial theorem, the kinetic energy term $K$ provides pressure support in halos, and can be separated into thermal and non-thermal components:
\begin{equation}
\label{eq:K}
\rho K = \frac{3}{2}\,P = \frac32(n\,k_B\,T + P_{\rm non-th})\;,
\end{equation}
where the relevant non-thermal motion is assumed to be isotropic on the halo scale, thus $2/3$ of the non-thermal (kinetic) energy contributes to pressure. 

We now introduce the reference potential energy for halos:
\begin{equation}
\label{eq:omega_grav_ref}
    \Omega_W^{\rm ref} = -\frac13\, f_{\rm b}\,\Omega_W^{\rm halo}\,,
\end{equation}
which corresponds to (minus) the amount of potential energy in halos contributing to the pressure. Here we have assumed that baryons and dark matter share the same kinetic energy per unit mass, therefore a baryon fraction $f_{\rm b}$ is needed in this expression. This may not be exact, as baryons and dark matter reach an equilibrium configuration in different ways. The former reaches equilibrium via collisions, whereas the latter is collisionless and thus achieves an equilibrium (virialized) state via exchange of energy between particles and time-dependent gravitational potential \citep[violent relaxation,][]{lynden-bell:1967,white:1996}. In addition, not all baryons remain in a hot gas phase, but some cool and form stars, making $f_{\rm gas}<f_{\rm b}$ \citep{bryan:2000}. While the gas fraction depends on halo mass, for the relevant cluster halos with $M_{500}>10^{14}~M_\sun$, $f_{\rm gas}/f_{\rm b} \sim 90$--95\% \citep[e.g.,][]{gonzalez/etal:2013}. We thus adopt $f_{\rm gas}/f_{\rm b}=0.925\pm 0.025$ and discuss the impact of this in the following estimate.

\begin{figure}[t]
 \includegraphics[width=8.5cm]{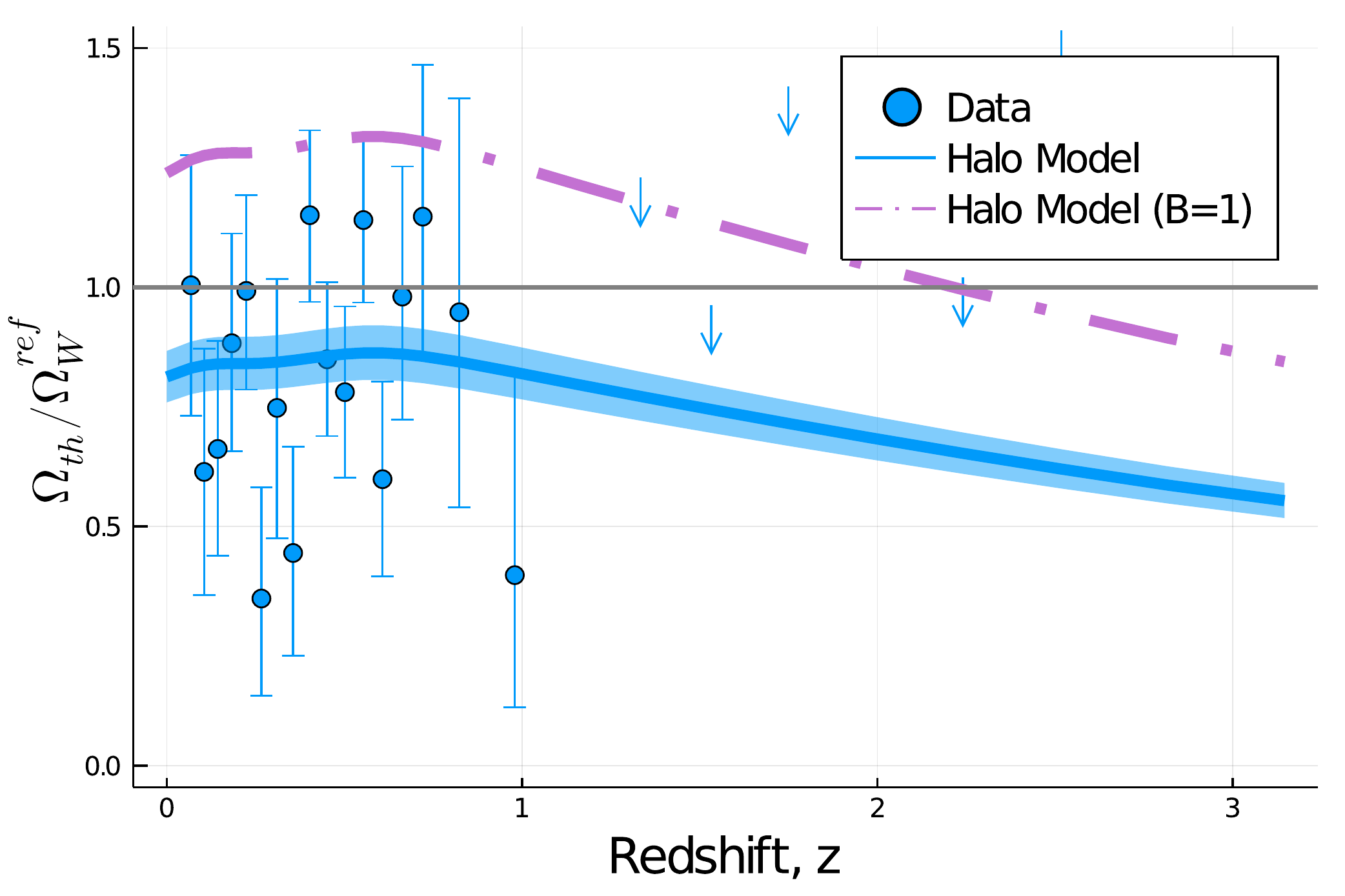}
\caption{
Efficiency of gravitational to thermal energy conversion given by the ratio of
the thermal energy density $\Omega_{\rm th}$ and the amount of potential energy in halos contributing to the pressure $\Omega_W^{\rm ref}$.
The data points with error bars and upper limits shown by the arrows are obtained by dividing the $\Omega_{\rm th}$ data shown in Figure~\ref{fig:omega_grav} by the best-fitting model for $\Omega_W^{\rm ref}$  given in Eq.~(\ref{eq:omega_grav_ref}). The model uncertainty is not included in the error bars.
The shaded area shows the best-fitting model with the 68\% confidence level
(the halo model for $\Omega_{\rm th}$ shown in Figure~\ref{fig:omega_grav} divided by the best-fitting model for $\Omega_W^{\rm ref}$.)
The dashed line shows the (non-physical) ratio for the model with no mass bias, $B=1$.
\label{fig:ratio}}
\end{figure}

Based on the assumptions used in this section, we can quantify the efficiency of gravitational to thermal energy conversion through the ratio
\begin{eqnarray}
\label{eq:eta}
    \frac{\Omega_{\rm th}}{\Omega_{W}^{\rm ref}} =
     \frac{1.7\pm 0.1}{2.07\pm 0.14}= 0.82\pm 0.07~~{\rm at~}z=0\,.
\end{eqnarray}
Our estimate indicates that, at the present time, the thermal energy accounts for about 80\% of the available reference gravitational energy. The redshift dependence of this ratio is shown in Figure~\ref{fig:ratio}. 
When correcting for $f_{\rm gas}/f_{\rm b}=0.925\pm 0.025$, the ratio becomes $0.89\pm 0.08$.

If we use the upper bound on the halo contribution (the upper boundary of the green shaded area in Figure~\ref{fig:omega_grav}), i.e., if we replace $\Omega_W^{\rm halo}$ by the difference between the non-linear and line power spectra, we find 
$\Omega_{\rm th}/\Omega_{W}^{\rm ref}\simeq 0.75$ (0.81 when including $f_{\rm gas}/f_{\rm b}=0.925$) for $0<z\le 0.5$ and 0.63 (0.68) for $0.5<z\le 1$.

We also find that, at $z<1$, this ratio is approximately 1.3 for the model with no mass bias, i.e. $B=1$. As $B$ is an empirical parameter for an empirical pressure profile model, $B=1$ does not represent a physical model in which the total pressure gradient balances gravity of an NFW profile, such as given in \citet{komatsu/seljak:2001}. 
Therefore, the $B=1$ model is not guaranteed to give $\Omega_{\rm th}/\Omega_{W}^{\rm ref}\le1$. Our approach could be used to put a lower limit on $B$ so that this ratio does not exceed unity.\\

Finally, we can quantify the non-thermal contribution to the total energy density. Expressing $P_{\rm non-th}$ given in Eq~(\ref{eq:K}) in terms of the density parameter of non-thermal pressure $\Omega_{\rm non-th}\equiv \Omega^{\rm ref}_W - \Omega_{\rm th}$, we get
\begin{equation}
\label{eq:non-thermal}
    \Omega_{\rm non-th} 
     =(0.37\pm 0.17)\times10^{-8}~~{\rm at~}z=0\;.
\end{equation}

\subsection{Comparison to the previous work}

In \citet{zhang/pen:2001}, the authors relate the gas temperature to the peculiar gravitational potential as $k_{\rm B}T_{\rm gas}=-[4m_{\rm p}/(8-5Y)](\phi-\bar\phi)/6$, where $\bar\phi({\bf x})=\int d^3x'\phi({\bf x'})W(|{\bf x}-{\bf x}'|)$ is a high-pass filtered potential. The gas density-weighted temperature, $\bar T_\rho$, is then given by the cross-correlation power spectrum of gas density and $\phi$. Using their Eq.~(6) in our Eq.~(\ref{eq:Trho}), we find
 \begin{equation}
     \Omega_{\rm th}^{\rm ZP01}
     = f_{\rm b}\frac{\Omega_{\rm m}^2H_0^2}{8\pi^2a}\int_0^\infty
     {\rm d}k~P_{\rm dg}(k)[1-W(k)]\,.
\end{equation}
The coefficient agrees with that of our $\Omega_W^{\rm ref}$. Here, $P_{\rm dg}(k)$ is the cross-correlation power spectrum of dark matter and gas, and $W(k)=\exp(-k^2r_{\rm e}^2/2)$ is the Fourier transform of the high-pass filter with a free parameter $r_{\rm e}$. The factor $1-W(k)$ acts to extract the small-scale contribution, which is dominated by the 1-halo term. Thus, their expression approximates $\Omega_W^{\rm ref}$. Our approach does not require $r_{\rm e}$ because we isolate the halo contribution using the 1-halo term or the difference between the linear and non-linear $P(k)$. 

The dark matter-gas cross-correlation power spectrum takes into account the difference in the spatial distribution of dark matter and gas on small scales. As the distribution of gas is smoother than that of dark matter on small scales, they write $P_{\rm dg}(k)=\exp(-k^2r_{\rm g}^2/2)P(k)$ with another free parameter $r_{\rm g}\sim 1/3~{\rm Mpc}/h$. This acts to reduce $\Omega_W^{\rm ref}$. We did not include this effect in our calculation, as this is subject to the effect of feedback by supernova explosion and AGN which are not well understood. This is included in the error budget we discussed at the end of Section~\ref{sec:accuracy}.

\subsection{Beyond virial theorem for the large-scale structure}

The virial relation, $K=-W/2$, does not hold for all the large-scale structure of the Universe. Here we need to take into account the fact that the systems of interest are not isolated and matter keeps accreting. 

The evolution of the kinetic and gravitational potential energies of collisionless non-relativistic particles is given by the Layzer-Irvine equation \citep{layzer:1963,irvine:1961,dmitriev/zeldovich:1963}
\begin{equation}
\label{eq:LI}
    \frac{\rm d}{{\rm d}t}(K+W)+\frac{\dot{a}}{a}(2K+W)=0\,,
\end{equation}
where the dot denotes the time derivative. The virial relation is the stationary solution of this equation.

\begin{figure}[t]
 \includegraphics[width=8.5cm]{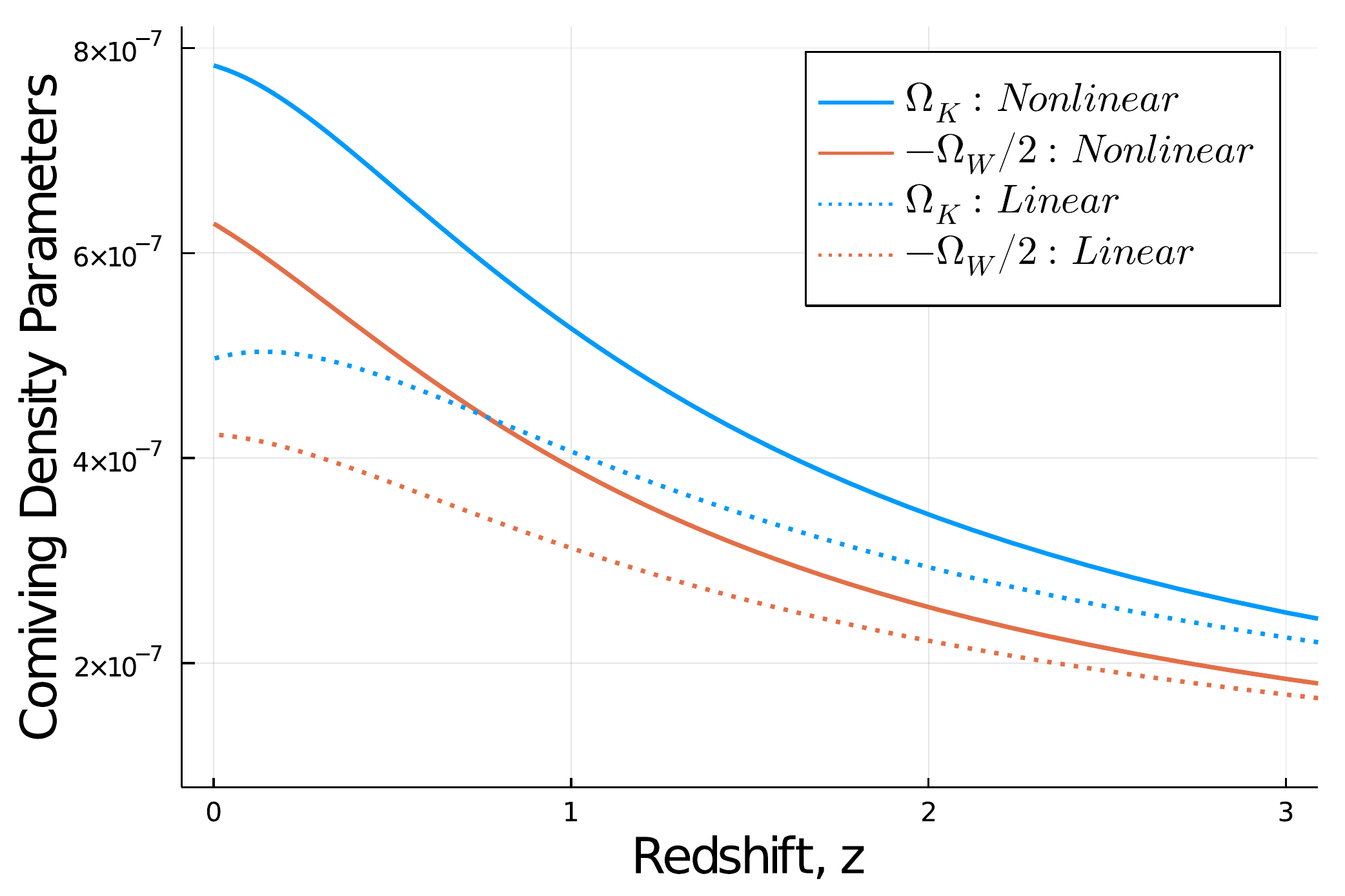}
\caption{
Solutions of the Layzer-Irvine equation~(\ref{eq:LI}). We show the comoving energy density parameters of $K$ ($\Omega_{K}$) and $-W/2$ ($-\Omega_{W}/2$). The horizontal axis shows the redshift, $z$. 
The solid and dotted lines show $\Omega_K$ (blue) and $\Omega_{W}$ (orange) for the non-linear and linear $P(k)$, respectively.
\label{fig:LI}}
\end{figure}

For linear theory \citep[see Eq.~(12) of][]{davis/miller/white:1997},
\begin{equation}
    K=-\frac{2f^2}{3\Omega_{\rm m}(a)}W\,,
\end{equation}
where $f\equiv {\rm d}\ln\delta_1/{\rm d}\ln a$ with the linear  density contrast $\delta_1$ and $\Omega_{\rm m}(a)=\Omega_{\rm m}/[a^3E^2(a)]$ is the matter density parameter at a given $a$. For the virial relation to hold, we need $f=\sqrt{3\Omega_{\rm m}/4}$ which is not satisfied in general.
The dotted lines in Figure~\ref{fig:LI} show the evolution of the comoving energy density parameters for $K$ (upper) and $-W/2$ (lower) computed from the linear $P(k)$. At $z=0$ the former is larger than the latter by a factor of 1.17.

To obtain $K$ for the non-linear $P(k)$, we solve the Layzer-Irvine equation numerically. We use the linear solution during the matter-dominated era, $K=-2W/3$, to set the initial condition. The solid lines in Figure~\ref{fig:LI} show the resulting comoving energy density parameters. At $z=0$ $K$ is larger than $-W/2$ by a factor of 1.25.\footnote{
One may wonder if we can use $W_{\rm 1h}(a)$ in Eq.~(\ref{eq:LI}) and solve for $K_{\rm 1h}(a)$. If we did this, we would find $K_{\rm 1h}> -W_{\rm 1h}/2$ by more than 40 percent despite that halos are virialized. However, this is not correct. Suppose that we split $W$ as $W=W_{\rm 1h}+W_{\rm other}$. While Eq.~(\ref{eq:LI}) is satisfied for the sum $W_{\rm 1h}+W_{\rm other}$, it is not satisfied for each component separately. Therefore, we only report $\Omega_{K}$ for the linear and non-linear $P(k)$. The linear $P(k)$ result is only an approximation and for comparison to the non-linear one.}
Thus, more kinetic energy is available in large-scale structure of the Universe than that of the stationary solution. This is due to the continuous infall of matter. 

The numerical values of the density parameter for the kinetic energy, $\Omega_K$, are given in Table~\ref{tab:kin}. These values should be compared with those in Table~\ref{tab:results}. 
This enhanced kinetic energy is associated with bulk flows of matter falling into the gravitational potential. During this stage, the gas carried with it has not yet thermalized with the surrounding hot gas. It is therefore not expected to contribute or perturb the global SZ signal, which is dominated by halo contributions. The solutions of the Layzer-Irvine equation found in this section are therefore not to be compared directly with $\Omega_{\rm th}$, but can be compared with the sum of the thermal SZ effect and, e.g., the kinetic SZ effect probing the bulk momentum density field.

\begin{table}
\centering
\caption{The density parameters of the kinetic energy in the large-scale structure of the Universe, $\Omega_{K}$, for the linear ($\Omega_{K}^{\rm lin}$) and non-linear $P(k)$ ($\Omega_{K}^{\rm nl}$).
The 1st column shows the redshift $z$, while the 2nd and 3rd columns show $10^7\Omega_{K}^{\rm lin}$ and $10^7\Omega_{K}^{\rm nl}$, respectively. The modelling uncertainties in $\Omega_{K}$ are equal to those in the corresponding $\Omega_{W}$ reported in Section~\ref{sec:accuracy}.
} \label{tab:kin}
\begin{tabular}{CCC}
\hline
\hline
z & 10^7\Omega_{K}^{\rm lin} & 10^7\Omega_{K}^{\rm nl} \\
\hline
0.0 & 4.97 & 7.83  \\
0.3 & 4.97 & 7.22  \\
0.5 & 4.76 & 6.65  \\
0.7 & 4.49 & 6.06  \\
1.0 & 4.07 & 5.26  \\
1.3 & 3.67 & 4.59  \\
1.5 & 3.43 & 4.21  \\
\hline
\end{tabular}
\end{table}

\section{Discussion and conclusion}\label{sec:conclusion}

We have presented estimates of the thermal energy density of baryons $\Omega_{\rm th}$ and the gravitational potential energy in collapsed halos $\Omega_{W}^{\rm halo}$. These quantities are related as the growth of structure leads to a conversion of gravitational energy associated with matter density fluctuations into kinetic energy, most of which thermalizes via shocks. 
Due to energy conservation, we expect $\Omega_{\rm th}/\Omega_{W}^{\rm ref}$ to be of order unity but, importantly, it should not exceed unity. Verifying this property is an interesting test of our understanding of baryons and dark matter as these two quantities are obtained using fundamentally different estimates: $\Omega_{W}^{\rm halo}$ is computed using the theoretical framework of the halo model which is driven by dark matter statistics, and $\Omega_{\rm th}$ using observations of the SZ effect.

Our analysis shows that $\Omega_{\rm th}/\Omega_{W}^{\rm ref}\simeq 0.8$ at $z=0$. It therefore provides an important check of our understanding of the related statistical properties of these quantities. Of particular interest, thanks to the enhanced accuracy of the $\Omega_{\rm th}$ and ${\Omega_{W}^{\rm ref}}$ estimates, we are now in a position to probe the amount of non-thermal pressure. Our analysis has allowed us to derive $\Omega_{\rm non-th} \simeq 0.37\times10^{-8}$ at $z=0$.

Now, what accounts for this? A promising candidate for non-thermal pressure quantified by $\Omega_{\rm non-th}$ is the bulk and turbulent motion of gas sourced by mass accretion and structure formation \citep{dolag/etal:2005,iapichino/niemeyer:2008,vazza/etal:2006,vazza/etal:2009,vazza/etal:2016,vazza/etal:2018,lau/kravtsov/nagai:2009,maier/etal:2009,shaw/etal:2010,iapichino/etal:2011,battaglia/etal:2012,nelson/etal:2014b,shi/komatsu:2014,shi/etal:2015,angelinelli/etal:2020}.
An analytical model \citep{shi/komatsu:2014} validated by cosmological hydrodynamical simulations \citep{shi/etal:2015} suggests the following picture: the non-thermal energy is sourced by the mass growth of halos via mergers and accretion, and dissipates with a time-scale determined by the turnover time of the largest turbulence eddies. The timescale of dissipation is on the order of the local dynamical time, which is shorter in the core and longer in the outskirts of halos. Thus, the fraction of the non-thermal energy relative to the total increases with radii, reaching tens of percent at the virial radius depending on halo mass and redshift. Therefore, the amount of non-thermal energy in this form seems sufficient to account for the non-thermal pressure budget given in Eq.~(\ref{eq:non-thermal}).

Our $\Omega_{\rm th}$ measurement, as well as the inferred amount of non-thermal pressure or energy density, should add to the list of useful, non-trivial tests for modern cosmological hydrodynamical simulations of formation and evolution of galaxies and clusters of galaxies \citep{vogelsberger/etal:2014,dubois/etal:2014,lebrun/etal:2014,schaye/etal:2015,dolag/komatsu/sunyaev:2016,mccarthy/etal:2017,pillepich/etal:2018}. 
It would provide further insights to the balance of thermal and non-thermal energy budgets of large-scale structure of the Universe, which is related to the physical ingredients of galaxy formation such as the AGN feedback and hot/cold mode gas accretion onto halos \citep{2005MNRAS.363....2K,2009Natur.457..451D}.

Observational evidence for non-thermal motion in galaxy clusters and groups includes line broadening of metal lines in X-ray bands \citep{sunyaev/norman/bryan:2003,inogamov/sunyaev:2003,hitomi/etal:2016,lau/etal:2017}, the kinetic SZ effect \citep{sunyaev/norman/bryan:2003,inogamov/sunyaev:2003,sayers/etal:2013,sayers/etal:2019,adam/etal:2017} and surface brightness fluctuations in X-ray \citep{schuecker/etal:2004,kawahara/etal:2008,churazov/etal:2012,churazov/etal:2016} as well as in the thermal SZ effect \citep{khatri/gaspari:2016,ueda/etal:2018,dimascolo/etal:2019}. On-going and future high sensitivity, high spectral and spatial resolution X-ray ({\sl eROSITA}, {\sl XRISM}, {\sl ATHENA})
and SZ experiments \citep[for a review]{mroczkowski/etal:2019} as well as cross-correlation of the future large-scale survey data of the microwave sky from Simons Observatory \citep{simonsobservatoru:2019} and CMB-S4 \citep{cmbs4:2016} with galaxy survey data \citep{battaglia/etal:2017,pandey/baxter/hill:2020}
would provide stringent tests of the thermal and non-thermal cosmic energy inventory.

\acknowledgments
EK thanks N. Afshordi, B. Bolliet, B. M. Sch\"afer, D. Spergel, and V. Springel for useful comments on the draft.  
This work was supported in part by NSF grant AST1313302 and NASA grant NNX16AF64G (YC, BM),
the Excellence Cluster ORIGINS which is funded by the Deutsche Forschungsgemeinschaft (DFG, German Research Foundation) under Germany's Excellence Strategy - EXC-2094 - 390783311 (EK), and JSPS KAKENHI Grant Number JP15H05896 (RM, EK) and JP20K14515 (RM). The Kavli IPMU is supported by World Premier International Research Center Initiative (WPI), MEXT, Japan. 

\bibliography{references}
\end{document}